%% file: main.tex
\documentclass[11pt]{article}

\usepackage[preprint]{acl}

\usepackage{times}
\usepackage{latexsym}

\usepackage[T1]{fontenc}

\usepackage[utf8]{inputenc}

\usepackage{microtype}

\usepackage{inconsolata}

\usepackage{graphicx}
\usepackage{booktabs}
\usepackage{multirow}
\usepackage{CJK}
\usepackage{algorithm}
\usepackage{algpseudocode}
\usepackage{booktabs}
\usepackage{multirow}
\usepackage{listings}
\usepackage{xcolor}
\usepackage{colortbl}
\usepackage[most]{tcolorbox}
\usepackage{enumitem}

\definecolor{promptbg}{RGB}{194,194,216}
\newtcolorbox{promptbox}[1][]{
    enhanced,
    colback=blue!7,
    colframe=blue!40!black,
    coltitle=white,
    fontupper=\small,
    fonttitle=\small,
    boxrule=0.5mm,
    arc=2mm,
    left=3mm, right=3mm, top=2mm, bottom=3mm,
    title={#1},
}

\definecolor{commentgreen}{RGB}{34, 139, 34}

\title{LTS-VoiceAgent: A Listen-Think-Speak Framework for Efficient Streaming Voice Interaction via Semantic Triggering and Incremental Reasoning}

\author{
 \textbf{Wenhao Zou\textsuperscript{1,2}}%
 \thanks{Work done during internship at Meituan.},
 \textbf{Yuwei Miao\textsuperscript{2}},
 \textbf{Zhanyu Ma\textsuperscript{1}},
 \textbf{Jun Xu\textsuperscript{1}}%
 \thanks{Corresponding authors.},
\\
 \textbf{Jiuchong Gao\textsuperscript{1}}\footnotemark[2],
 \textbf{Jinghua Hao\textsuperscript{1}},
 \textbf{Renqing He\textsuperscript{1}},
 \textbf{Jingwen Xu\textsuperscript{1}}
\\
\\
 \textsuperscript{1}Meituan,
 \textsuperscript{2}University of Chinese Academy of Sciences
}

\begin{document}
\maketitle
\begin{abstract}
    Real-time voice agents face a dilemma: end-to-end models often lack deep reasoning, while cascaded pipelines incur high latency by executing ASR, LLM reasoning, and TTS strictly in sequence, unlike human conversation where listeners often start thinking before the speaker finishes. Since cascaded architectures remain the dominant choice for complex tasks, existing cascaded streaming strategies attempt to reduce this latency via mechanical segmentation (e.g., fixed chunks, VAD-based splitting) or speculative generation, but they frequently either break semantic units or waste computation on predictions that must be rolled back. To address these challenges, we propose \textbf{LTS-VoiceAgent}, a \textbf{Listen–Think–Speak} framework that explicitly separates \emph{when to think} from \emph{how to reason incrementally}. It features a Dynamic Semantic Trigger to detect meaningful prefixes, and a Dual-Role Stream Orchestrator that coordinates a background \textbf{Thinker} (for state maintenance) and a foreground \textbf{Speaker} (for speculative solving). This parallel design enables ``thinking while speaking'' without blocking responses. We also introduce a Pause-and-Repair benchmark containing natural disfluencies to stress-test streaming robustness. Experiments across VERA, Spoken-MQA, BigBenchAudio, and our benchmark show that LTS-VoiceAgent achieves a stronger accuracy–latency–efficiency trade-off than serial cascaded baselines and existing streaming strategies.
\end{abstract}

\section{Introduction}

\begin{figure}[t]
  \centering
  \includegraphics[width=0.92\columnwidth]{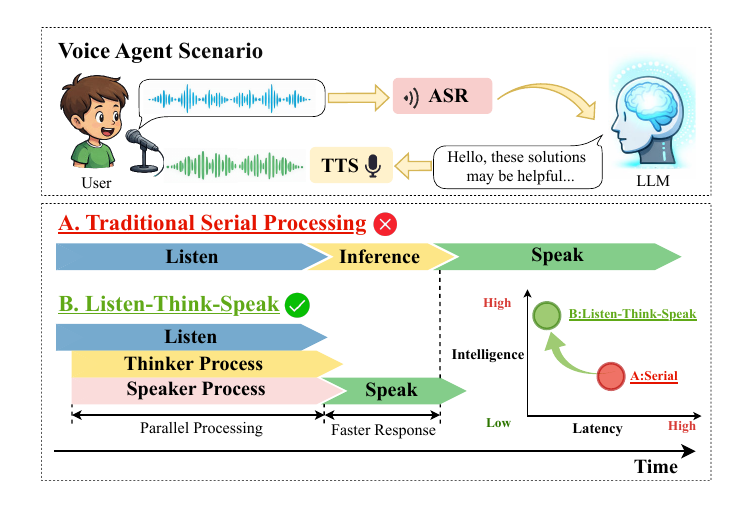}
  \caption{Traditional Serial Processing vs. LTS-VoiceAgent Framework for Voice Interaction.}
  \label{fig:profile}
  \vspace{-0.5em}
\end{figure}

\begin{figure*}[tbp]
  \includegraphics[width=1\linewidth]{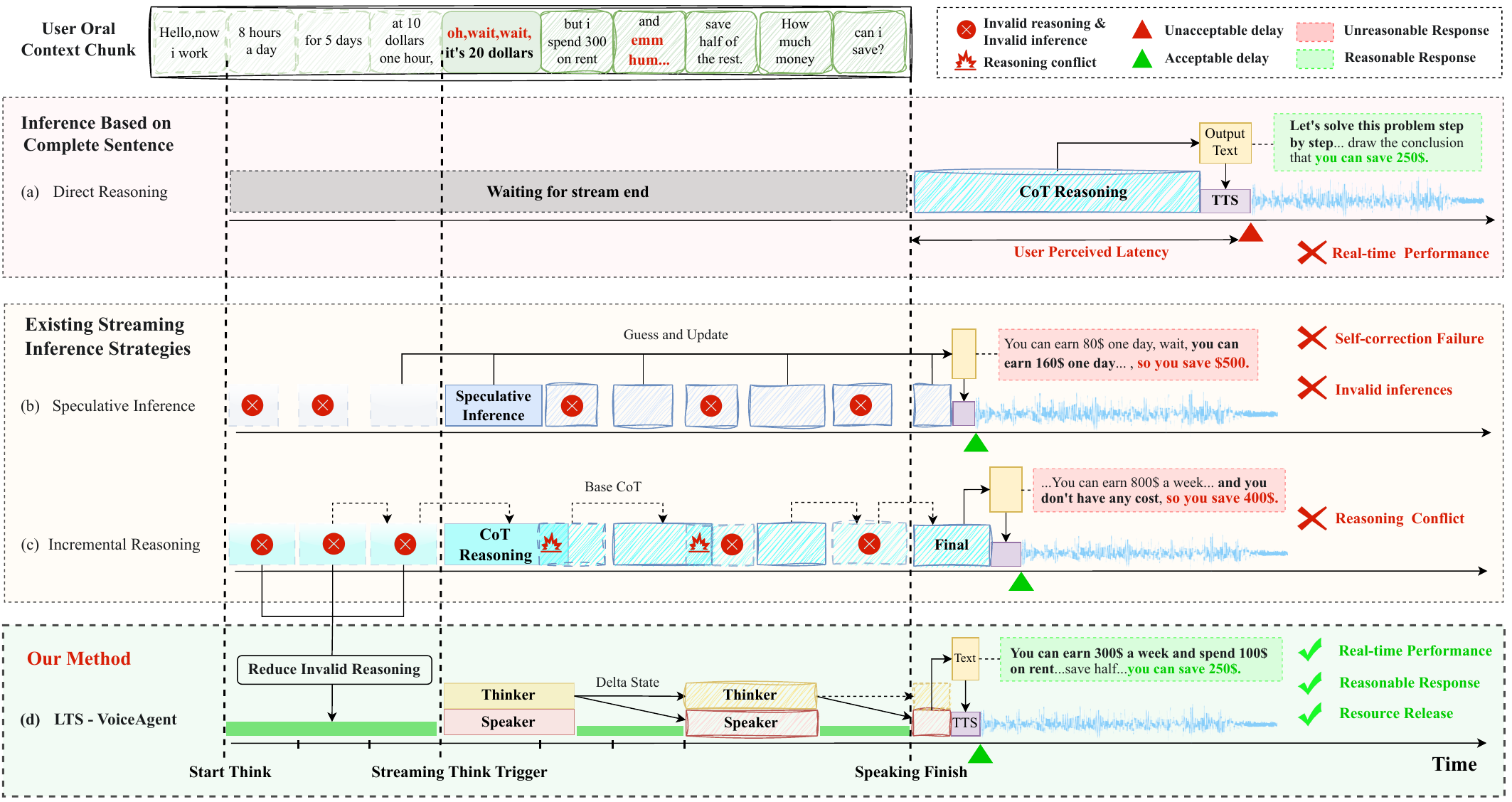}
  \caption{Examples of different reasoning methods.}
  \label{fig:motivation}
\end{figure*}

The convergence of reasoning-capable Large Language Models \citep{brown2020language, achiam2023gpt, wei2022chain} and high-fidelity speech foundation models \citep{radford2023robust, wang2023neural} has catalyzed a transformative shift in intelligent systems. Consequently, Voice AI is rapidly evolving from simple command-execution bots to sophisticated conversational agents capable of handling complex, long-horizon tasks \citep{hurst2024gpt, huang2024audiogpt}. In these scenarios, user expectations are twofold and often conflicting: they demand the \textbf{instant responsiveness} of a human listener (i.e., low latency) \citep{levinson2016turn} and the \textbf{complex reasoning capabilities} of a domain expert \citep{zhang2023speechgpt}.

While End-to-End (E2E) models reduce latency \citep{fang2024llama, defossez2024moshi}, they often lag in reasoning capabilities, making cascaded architectures the dominant solution for complex agentic tasks \citep{chen2024voicebench, jain2025voiceagentbench}. However, this dominance comes with a latency cost. As contrasted in Figure \ref{fig:profile}, traditional serial pipelines (Figure \ref{fig:profile}A) prioritize accuracy but suffer from significant delays, whereas our LTS framework (Figure \ref{fig:profile}B) aims to achieve fast responsiveness within this high-intelligence architecture. Further mechanism analysis in Figure \ref{fig:motivation}(a) reveals the root cause: serial systems are architecturally bound to silence while the user speaks, disrupting natural turn-taking \citep{skantze2021turn}. To mitigate this, existing streaming strategies attempt to introduce parallelism, yet they introduce new challenges:

\begin{itemize}
    \item \textbf{Inefficient Speculative Inference:} Some methods attempt to "guess" the user's intent before they finish speaking \citep{li2025predgen}. As illustrated in \textbf{Figure \ref{fig:motivation}(b)}, this leads to significant waste when the user's intent shifts (e.g., "oh, wait..."). The model must frequently rollback invalid predictions, resulting in a "Self-correction Failure."
    \item \textbf{Fragmented Incremental Reasoning:} Simply activating reasoning on streaming chunks leads to the chaotic state shown in \textbf{Figure \ref{fig:motivation}(c)}. This high-frequency triggering generates massive "Invalid Thinking" and Reasoning Conflicts, as the model forces inferences on semantically incomplete fragments. Even when heuristic constraints like Voice Activity Detection (VAD) \citep{chen2024livemind} are applied, the fundamental limitation persists: since VAD relies on acoustic rules (e.g., silence duration) rather than semantic completeness, it inevitably triggers reasoning during meaningless hesitations, perpetuating the issues of computational waste and logical inconsistency.
\end{itemize}

To address these challenges—specifically identifying \textit{when} to think and \textit{how} to reason incrementally over dynamic streams—we propose \textbf{LTS-VoiceAgent}, a \textbf{L}isten-\textbf{T}hink-\textbf{S}peak framework designed for efficient streaming interaction. As depicted in \textbf{Figure \ref{fig:motivation}(d)}, our approach differs fundamentally from prior paradigms: instead of passive waiting or aggressive guessing, LTS-VoiceAgent employs a \textbf{Dynamic Semantic Trigger} to filter out non-informative fragments (e.g., hesitation markers like "hum...") and activates reasoning only when semantic information is sufficient. Furthermore, to ``process the growing context'' without being misled by disfluencies, we devise a collaborative reasoning mechanism governed by a Stream Orchestrator. Instead of treating updates as independent queries, this design maintains a continuous reasoning chain where a background \textbf{Thinker} iteratively refines the mental state and passes the evolved context to a foreground \textbf{Speaker}. This allows the system to reuse prior deductions and self-correct during pauses, ultimately enabling instant Answer-first'' responses once the user finishes speaking.

In summary, our contributions are as follows:
\begin{itemize}
    \item \textbf{Dynamic Semantic Trigger:} We propose a lightweight module that identifies valid semantic boundaries within the continuous stream. By detecting when the input context is semantically sufficient, it triggers reasoning at the optimal moment, effectively avoiding redundant computation caused by premature execution.
    \item \textbf{Dual-Role Stream Orchestrator:} We design a collaborative inference architecture that leverages batch processing to parallelize understanding and generation. By coordinating a \textbf{Thinker} for robust state tracking (input sanitization and planning) and a \textbf{Speaker} for speculative execution, this mechanism enables the model to handle ``pause-and-repair'' behaviors and intent drifts effectively while achieving millisecond-level latency.
    \item \textbf{Pause-and-Repair Benchmark:} To evaluate streaming agents in realistic conditions, we construct a dataset based on GSM8K and MMLU-PRO, enriched with natural speech phenomena such as hesitations, self-corrections, and varying speaking rates. Experiments show that LTS-VoiceAgent achieves superior performance in latency and quality while reducing inference costs compared to strong baselines.
\end{itemize}

\section{Related Work}

Research on real-time spoken interaction is rapidly shifting from traditional ``instruction--response'' systems to full-duplex conversational agents. Existing work can be broadly grouped into three lines.

\subsection{End-to-End Omni-Modal Models}

End-to-end models aim to eliminate the latency of intermediate text transcription by directly mapping audio features to semantic representations and capturing paralinguistic cues. Early work such as \citet{ren2020simulspeech} introduced wait-k style simultaneous decoding, while more recent systems \citep{fang2024llama, fang2025llama} achieve parallel speech decoding without waiting for ASR. To improve alignment, \citet{wang2024freeze} aligns the speech encoder with a frozen LLM, and \citet{zhang2024intrinsicvoice,long2025vita,shi2025voila} explore ``speech-in, speech-out'' models that better preserve prosody and emotion. However, these omni-modal models still lag behind large text-only LLMs in complex reasoning and controllable instruction following, which limits their use in high-stakes agentic tasks.

\subsection{Advanced Cascaded Architectures: Component Acceleration}

Cascaded architectures (ASR--LLM--TTS) remain dominant for applications that require strong reasoning and controllability, since they can plug in SOTA components from different domains. Recent work focuses on aggressively accelerating ASR and TTS to approach the latency of end-to-end systems: \citet{wei2025specasr} apply speculative decoding with a draft model for ASR, while \citet{du2024cosyvoice,shikhar2025llmvox} push streaming TTS towards sub-200ms latency and flexible style control. Architectural variants such as Audio-LLM front-ends \citep{li2024style} further improve expressiveness. However, most of these efforts optimize single components and still follow a serial pipeline where the LLM only starts reasoning after receiving a complete transcript.

\begin{figure*}[tbp]
  \includegraphics[width=0.99\linewidth]{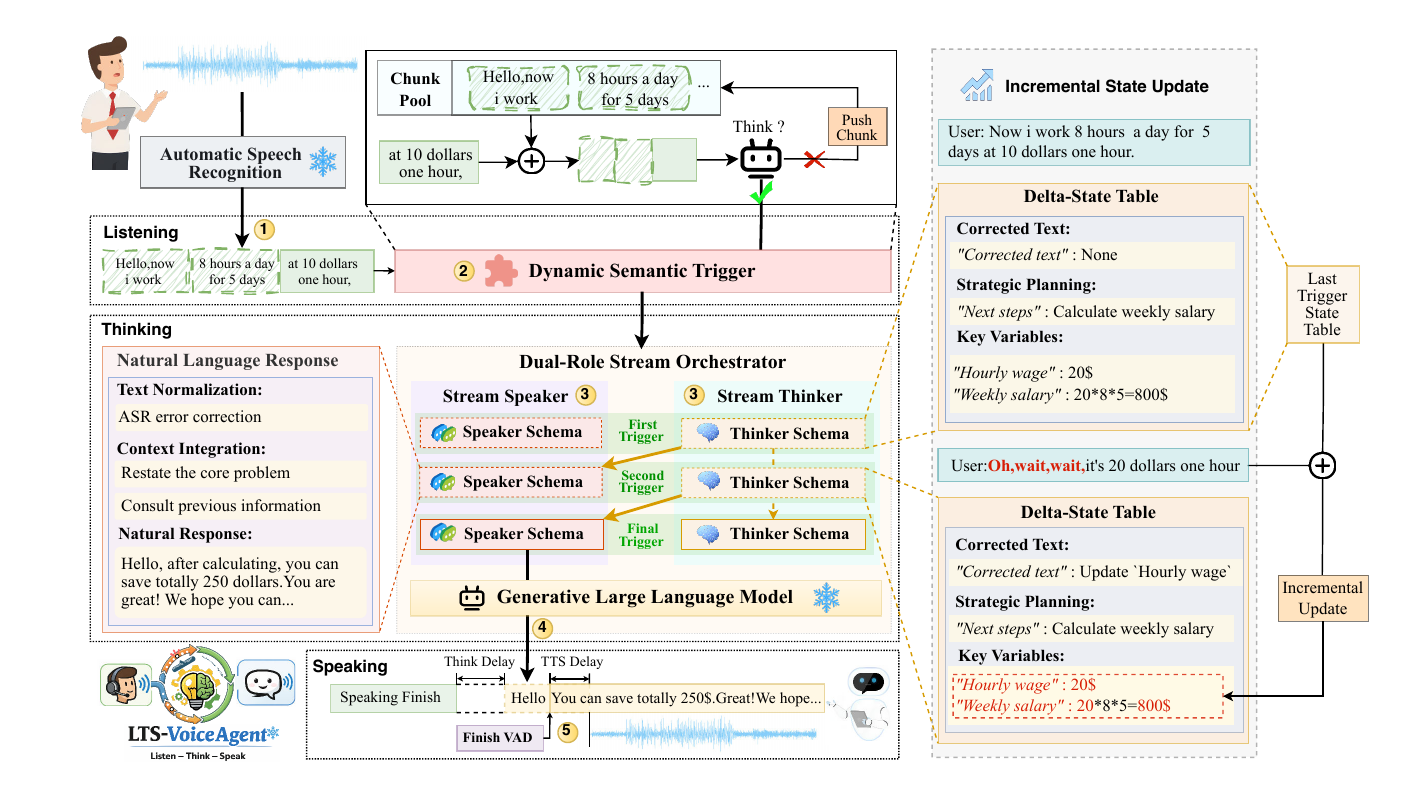}
  \caption{Overview of LTS-VoiceAgent.}
  \label{fig:framework}
\end{figure*}
\subsection{Streaming Inference and Speculative Strategies}

To break the limitations of serial waiting, recent work explores ``thinking while listening'' strategies that trade off predictive generation and computational cost. \citet{li2025predgen} propose a listen-and-guess approach that pre-generates responses from incomplete prompts, but frequent rollbacks lead to substantial wasted computation. An alternative line is segmented reasoning and context reuse: \citet{chen2024livemind} split the stream by clauses and store intermediate states in memory, and engineering systems such as \citet{likhomanenko2025chipchat,ethiraj2025toward} adopt sentence-level streaming and multi-threaded concurrency. However, their reliance on mechanical segmentation (punctuation or fixed length) often breaks semantic units. To improve accuracy, \citet{goel2025audio,woo2025think} introduce Chain-of-Thought (CoT) into spoken interaction at the cost of higher latency, while general-domain approaches \citep{cheng2024compressed,xu2025chain,hou2025thinkprune} accelerate CoT via token compression or pruning but assume static, complete inputs and lack mechanisms for state retention and reuse in dynamic streaming contexts.

In summary, this work addresses the critical latency--intelligence trade-off in voice agents, aiming to reconcile the responsiveness of end-to-end interactions with the reasoning depth of cascaded architectures. We propose LTS-VoiceAgent, which replaces mechanical segmentation with a Dynamic Semantic Trigger for precise activation and employs a Dual-Role Stream Orchestrator to coordinate background state maintenance (\emph{Thinker}) with foreground speculative generation (\emph{Speaker}). This design enables ``thinking while listening,'' achieving high-performance reasoning with millisecond-level latency.

\section{Methodology}
\label{sec:method}

This section details \textbf{LTS-VoiceAgent}, a unified framework designed to address the challenges of ``when to think'' and ``how to reason incrementally'' in streaming voice interactions. As illustrated in Figure \ref{fig:framework}, the system consists of two core components: (1) a \textbf{Dynamic Semantic Trigger}, which detects reasoning-worthy semantic boundaries in real-time streaming text; and (2) a \textbf{Reasoning State Orchestrator}, which maintains a structured \textbf{Delta-State CoT} to enable instant user response and asynchronous background state updates.

\subsection{Overview}
In standard cascaded voice interaction, the text stream output by the ASR, denoted as $S=\{s_1, s_2, ..., s_t\}$, grows continuously. Traditional strategies often fall into two extremes: either relying on acoustic VAD, which leads to frequent interruptions by meaningless hesitations (e.g., ``um...''), or employing aggressive speculative generation that results in computational waste. LTS-VoiceAgent proposes a ``Listen-Think-Speak'' paradigm based on semantic awareness.

The system workflow is as follows: Text chunks incrementally output by the ASR are first processed by the \textbf{Semantic Trigger}. This module calculates the semantic saturation probability $P_{trigger}$ of the current prefix $S_{1:t}$. Only when $P_{trigger} > \tau$ does the system activate the LLM for reasoning. Consequently, the model performs two types of reasoning: (1) \textit{thinking while listening}, where the model updates internal states during the user's speech; and (2) \textit{direct responding}, where the model generates an answer. To address overlap and correction issues in streaming input, the system continuously maintains a reasoning state table that preserves user intent and intermediate deductions. Once the user finishes speaking, the final response is generated directly based on this accumulated state table.

\subsection{Dynamic Semantic Trigger}
The core objective of the trigger is to allocate computational costs to moments characterized by a ``significant information increment'' and ``sufficient time for reasoning,'' rather than mere acoustic silence. We model this as a streaming binary classification task.

\subsubsection{Data Synthesis via Semantic Alignment}
Due to the scarcity of large-scale annotated ``semantic breakpoints,'' we designed a self-supervised data synthesis pipeline using LLMs. \textbf{We select raw texts $D_{raw}$ from representative reasoning-intensive corpora to serve as the seed data.} This data-agnostic design allows the framework to adapt to various domains (e.g., mathematics, coding, or general knowledge) by simply switching the source texts.

We employ GPT-4o as a ``linguistic analysis expert'' to insert special trigger tokens ``[T]'' into the text. The insertion rules adhere to semantic completeness principles: tokens appear only at the boundaries of clauses, constraints, or complete intent expressions, specifically where no massive information influx is expected immediately, while strictly skipping disfluent hesitation markers. Following automated annotation, the authors manually verified the dataset to ensure the validity of the semantic boundaries. Subsequently, we simulate the ASR transmission process at varying speaking rates, slicing the verified text containing ``[T]'' into streaming prefix sequences $x_{1:k}$. For each prefix, if its end crosses a ``[T]'' token, it is labeled as a positive instance $y=1$; otherwise, it is negative $y=0$. This process constructs a binary classification dataset $D_{trigger}$ optimized for streaming interaction.

\subsubsection{Trigger Inference}
We employ a lightweight DistilBERT as the backbone for the trigger. During inference, the trigger encodes the accumulated transcript from the ASR and outputs a confidence score. To prevent jitter, we implement a dual-filtering mechanism: (1) \textbf{Threshold Gating:} Triggering occurs only when the score exceeds $\tau$; (2) \textbf{Deduplication Suppression:} If the current text is identical to the text at the last trigger (indicating the ASR is in a non-steady jitter state), triggering is suppressed. This design ensures that the system consumes reasoning resources only when the user expresses new valid information.

\subsection{Dual-Role Stream Orchestrator}

To reconcile the conflict between deep semantic understanding and millisecond-level responsiveness in streaming interaction, we propose the \textbf{Dual-Role Stream Orchestrator}. Responding within a single inference pass often leads to frequent interruptions and semantic fragmentation (or ``thought discontinuity''), even with trigger controls. Therefore, unlike traditional approaches that couple understanding and generation into a single serial process, we leverage the \textbf{batch processing capabilities of the inference engine} to concurrently schedule two complementary roles: the \textbf{Thinker} and the \textbf{Speaker}. Through an asynchronous communication mechanism, the orchestrator enables the Speaker to perform \textbf{Speculative Solving} before the Voice Activity Detection (VAD) finalizes the turn, while utilizing the global state maintained by the Thinker in the background to guide the logical direction of the generation.

\subsubsection{Thinker: Stream Analysis \& State Maintenance}
The Thinker functions as a background analysis module dedicated to maintaining a structured understanding of user intent without directly participating in response generation. Addressing the noise inherent in streaming ASR, the Thinker first performs \textbf{Input Sanitization}, utilizing context to correct phonetic errors in the transcript (e.g., correcting ``sign'' to ``sine''). Subsequently, it evaluates semantic completeness, extracts \textbf{Key Variables}, and formulates a dynamic \textbf{Plan} for the next execution step. The output of the Thinker is encapsulated into a compact JSON \textbf{State Snapshot}, comprising the following core fields:
\begin{itemize}
    \item \texttt{corrected\_text}: The semantically sanitized text serving as the system's internal ground truth.
    \item \texttt{key\_variables}: Extracted task-relevant entities or parameters.
    \item \texttt{plan}: High-level logical steps guiding subsequent reasoning.
\end{itemize}

\subsubsection{Speaker: Speculative Execution}
The Speaker acts as the foreground executor responsible for generating the final user-facing response. To overcome the latency bottleneck caused by serial waiting, the Speaker adopts an \textbf{Answer-First} strategy similar to PredGen: reasoning is triggered immediately upon activation, without waiting for the user's speech to fully cease. Its inference process follows a ``\textbf{Restate-Consult-Solve}'' paradigm: First, the Speaker briefly restates the core problem for implicit self-verification (\textbf{Restate \& Verify}). Next, it consults the decision plan in the shared memory (\textbf{Consult Plan}); if the plan is clear, the Speaker follows its steps, otherwise, it constructs logic based on the raw context. Finally, it outputs the specific response.

\subsubsection{Orchestrator: Asynchronous Coordination Mechanism}
The orchestrator manages the asynchronous collaboration and interruption handling between the Thinker and the Speaker. Regarding communication, the system employs a \textbf{State Injection} mechanism: the \texttt{plan} and \texttt{key\_variables} generated by the Thinker in the \textit{previous} turn are injected into the \textit{current} Speaker's context as prior knowledge. This design enables the Speaker to ``stand on the shoulders of the Thinker,'' benefiting from deep planning without incurring the latency penalty of waiting for the Thinker's real-time inference. Regarding interruption handling, when a substantial semantic shift is detected in the user's input stream, the orchestrator immediately terminates the current Speaker process but preserves the valid state generated by the Thinker. This state is then merged into the next context, achieving efficient ``thought rollback'' and state updates.

\section{Dataset: A Pause-and-Repair Benchmark}
\label{sec:dataset}

To bridge the gap between static text benchmarks (e.g., GSM8K\citep{cobbe2021training}, MMLU\citep{hendrycks2020measuring}) and real-world spoken interaction, and to verify the robustness of our framework in noisy scenarios, we constructed a large-scale synthetic streaming speech dataset. This dataset simulates the natural flux of human speech, providing a rigorous and realistic testbed for next-generation voice agents.

\paragraph{Scenario-based Contextualization}
To ensure the model triggers deep reasoning rather than simple pattern matching, we selected \textbf{GSM8K} and \textbf{MMLU-Pro}\citep{wang2024mmlu} as source data. We implemented a \textbf{Scenario-based Contextualization} strategy to transform abstract exam questions into concrete, first-person help-seeking scenarios (e.g., homework tutoring). This processing ensures the \textbf{semantic self-containment} of the audio input, compelling the model to perform end-to-end deep reasoning based on a single audio stream without relying on external text prompts.

\input{table_acc_latency.tex}

\paragraph{Hybrid Spontaneous Simulation}
Given the high disfluency of authentic spoken language, we designed a \textbf{Hybrid Linguistic Perturbation} pipeline (see Appendix \ref{sec:appendix_data}) to simulate "thinking-while-speaking" behavior. Leveraging the reasoning capabilities of the GPT-4o, we constructed a challenging \textbf{"Hybrid"} evaluation set.

Instead of simple noise injection, this dataset simultaneously integrates two complex speech characteristics:
\begin{itemize}
    \item \textbf{Filler Words Injection:} We introduced a natural hesitation mechanism that dynamically controls the density of fillers (e.g., ``um,'' ``uh'') and discourse markers based on text length, carefully avoiding semantic collapse caused by over-padding.    
    \item \textbf{Self-Correction Patterns:} Addressing dynamic intent shifts, we designed four self-correction paradigms, including entity correction and intent reset. We utilized Instruction Tuning to ensure that these correction behaviors align with human cognitive habits.
\end{itemize}

Finally, all text variants were converted into waveform data via a high-quality open-source \textbf{TTS component \citep{du2024cosyvoice}}, prioritizing the replication of the temporal characteristics and logical discontinuities inherent in human conversation.

\section{Experiments}
\label{sec:experiments}

In this section, we evaluate LTS-VoiceAgent in terms of reasoning quality, response latency, and inference efficiency.

\subsection{Experimental Setup}

\subsubsection{Datasets}
We evaluate LTS-VoiceAgent on three representative public benchmarks and our synthetic dataset. \textbf{In contrast to closely related setups that approximate streaming ASR by feeding fixed text chunks or injecting heuristic noise to mimic recognition errors and revisions, we adopt a audio-only evaluation pipeline that exposes the system to realistic ASR behavior, including misrecognitions, non-uniform speaking rates, and related artifacts.} Specifically, we use \textbf{VERA} (AIME and GPQA-Diamond) for challenging math and science reasoning \citep{lin2025voice}, \textbf{Spoken-MQA} for multi-step spoken math QA \citep{wei2025towards}, \textbf{BigBenchAudio} for diverse open-domain spoken tasks \citep{suzgun2022challenging}, and our \textbf{Pause-and-Repair Benchmark} (Sec. \ref{sec:dataset}) featuring extensive hesitations and self-corrections.

\subsubsection{Baselines}
We compare LTS-VoiceAgent against three streaming paradigms: \textbf{Serial Reasoning (No-think / Think)}, which runs the LLM only after ASR completes and provides an accuracy upper bound; \textbf{PredGen}, which triggers chunk-based speculative generation on fixed text segments \citep{li2025predgen}; and \textbf{LTS-VAD}, which augments PredGen with voice activity detection so that reasoning is triggered mainly during silences. 
In addition, we compare against an end-to-end speech model, Qwen2.5-Omni \citep{xu2025qwen2}, to quantify how cascaded pipelines differ from end-to-end modeling in the accuracy--latency trade-off.

\subsubsection{Implementation \& Metrics}
We control for ASR variability by using a unified internal streaming ASR API across all experiments, and we fix the LLM backbone to \textbf{Qwen3-8B}. All experiments are executed on a single NVIDIA A100 GPU, and all baselines are evaluated under the same hardware and ASR setup for fairness. We implement LTS-VoiceAgent with an asynchronous dual-stream inference stack based on vLLM. The streaming ASR transcript is processed at a 200ms granularity (16kHz, mono PCM; 3,200 samples per chunk). For the Dynamic Semantic Trigger, we use a lightweight DistilBERT classifier with maximum input length 512 and threshold gating at $\tau=0.65$, together with deduplication suppression that only prevents repeated triggers on identical transcripts under ASR jitter. 

\noindent\textbf{Trigger Implementation Details.}
We instantiated the training set with \textbf{1000 samples each from GSM8K and MMLU-Pro}. The streaming decomposition (Sec. 3.2.1) expanded these texts into cumulative prefixes, creating a large-scale sample pool. After applying \textbf{1:1 negative downsampling} to balance the dataset, we obtained \textbf{100k samples} to fine-tune the DistilBERT classifier (learning rate 5e-5). We evaluate the trigger's effectiveness implicitly through end-to-end system metrics rather than standalone classification accuracy, acknowledging the non-unique nature of semantic boundaries in spoken language.

We evaluate \textbf{quality} via Accuracy, \textbf{latency} via Time to First Token and Time-to-First-Sentence (TTFS), and \textbf{efficiency} via the Number of Forward-pass Evaluations (NFE) and the Number of Interruption Times (NIT). To quantify trigger validity, we report the \textbf{Interruption Rate}, defined as $\text{NIT}/\text{NFE}$; lower values indicate fewer mid-reasoning interruptions caused by subsequent speech and therefore more precise trigger timing. Unless otherwise noted, both the Speaker and Thinker streams use greedy decoding with temperature $0$ and a maximum of 4,096 new tokens.
\subsection{Experimental Results}

\input{table_efficiency.tex}

\begin{figure}[tbp]
    \centering
    \includegraphics[width=\linewidth]{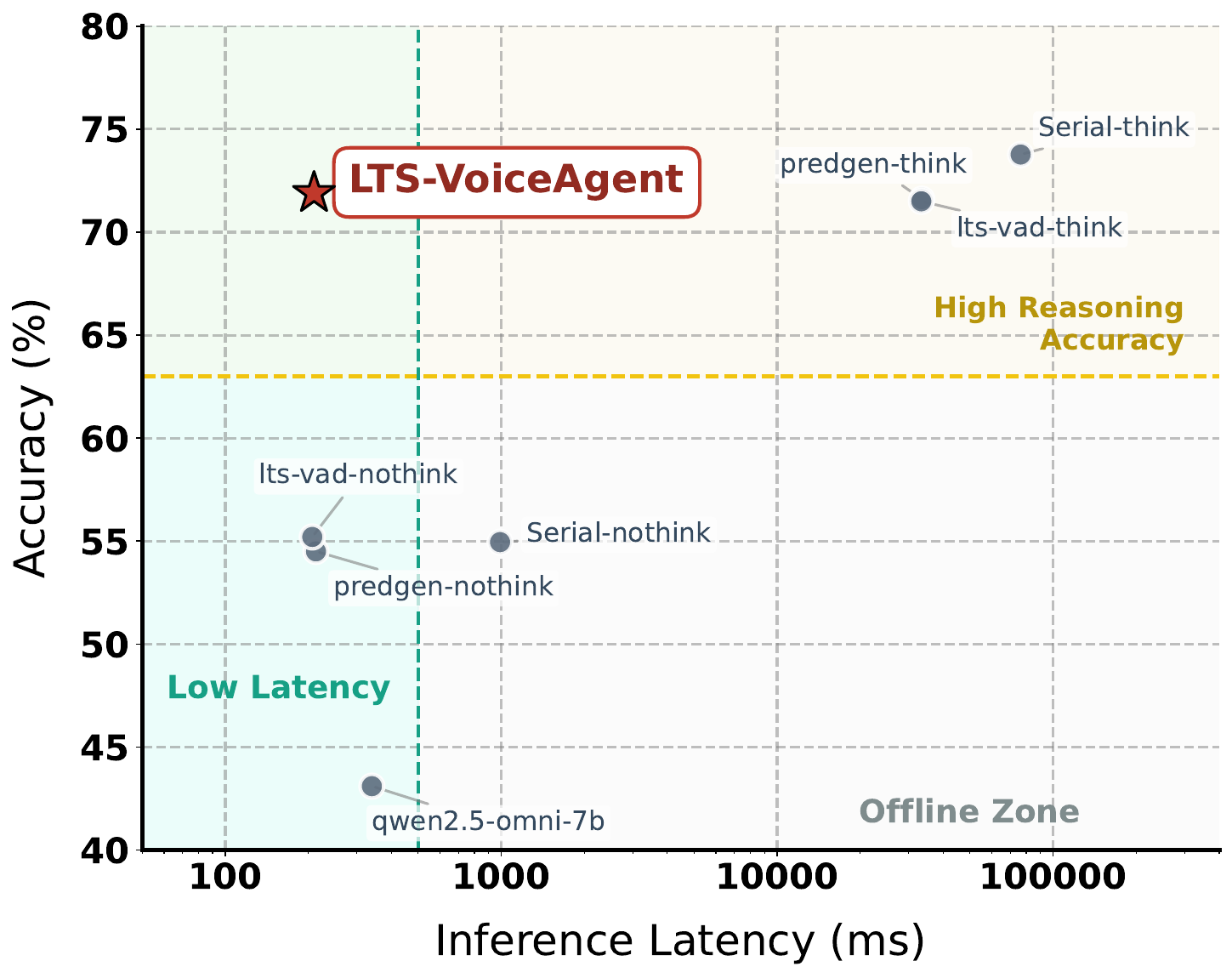}
    \caption{Performance (Accuracy) vs. Latency trade-off analysis.}
    \label{fig:bigbench_trade_off}
\end{figure}

Table \ref{tab:main_results_consolidated} presents comprehensive results on the three public datasets. To better visualize the quality--latency trade-off, we plot Accuracy against latency for \textbf{BigBenchAudio} by converting the corresponding results into a 2-D scatter plot, as shown in Figure \ref{fig:bigbench_trade_off}.

\textbf{LTS-VoiceAgent offers the best overall trade-off.} Table~\ref{tab:main_results_consolidated} shows that LTS-VoiceAgent matches streaming-level latency while achieving markedly higher accuracy than chunk- and VAD-based baselines across all benchmarks. Its TTFS remains sub-second, with only a small increase over the most aggressive chunking variants.

\textbf{Serial (Think)} yields the highest accuracy with full context, but incurs TTFS in the tens to hundreds of seconds, making it impractical for interactive agents. LTS-VoiceAgent narrows this quality gap while retaining real-time responsiveness.

To assess \textbf{efficiency}, Table \ref{tab:efficiency_consolidated} reports NFE, NIT, and the interruption rate. \textbf{PredGen} is highly wasteful: it performs frequent speculative evaluations and almost all of them are interrupted, indicating pervasive rollback/recompute. \textbf{LTS-VAD} reduces the evaluation count but still incurs substantial interruptions, with rates ranging from 41\% to 90\% depending on the dataset and mode. In contrast, \textbf{LTS-VoiceAgent} consistently achieves the lowest overhead and \textbf{keeps} interruptions near zero, resulting in \textbf{single-digit interruption rates} across all benchmarks. Trigger inference is lightweight, with an average runtime of about \textbf{5\,ms.}

\begin{figure}[tbp]
    \centering
    \includegraphics[width=\linewidth]{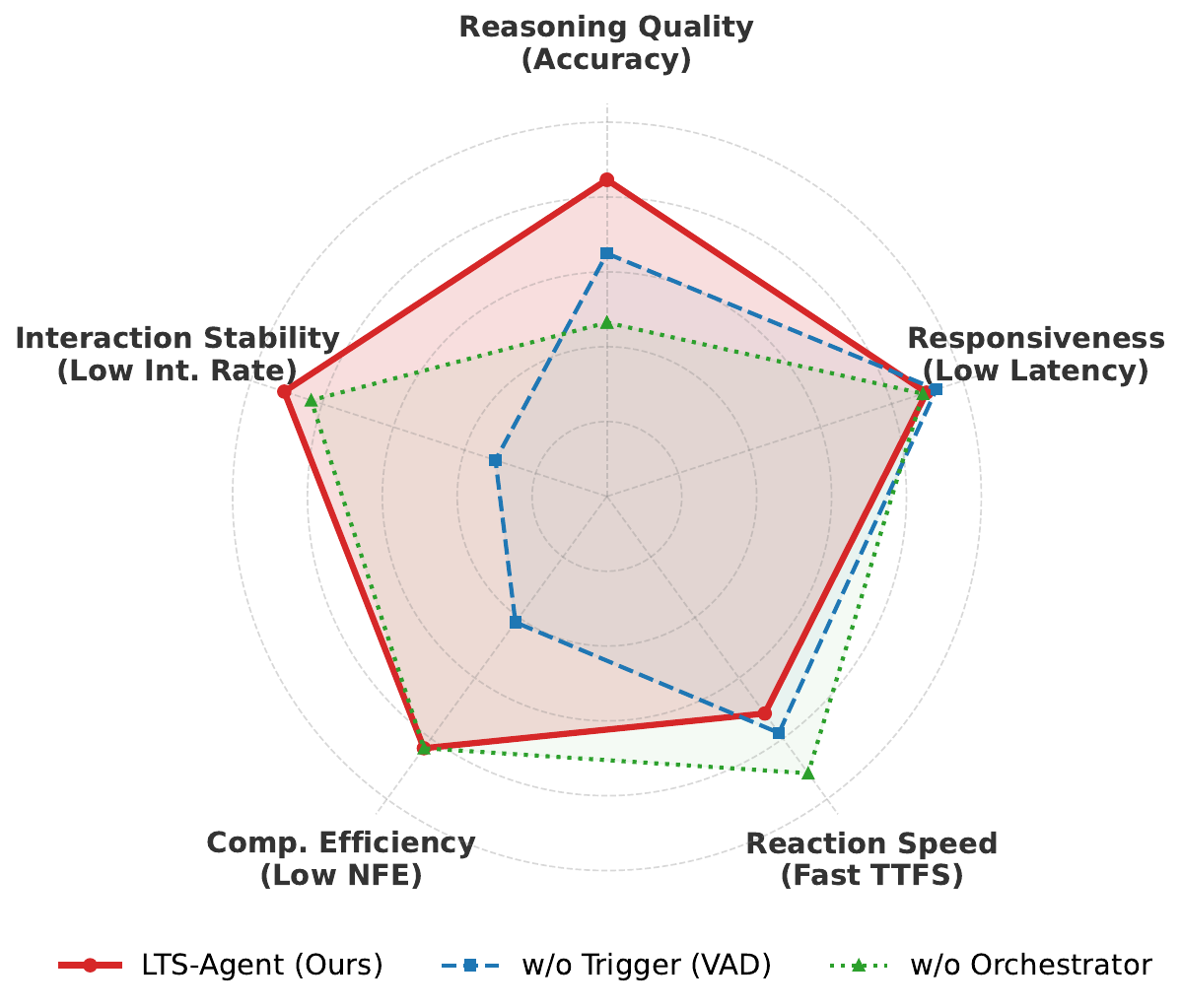}
    \caption{Ablation study on the LTS-VoiceAgent.}
    \label{fig:ablation_radar}
\end{figure}

\input{table_ablation_avg.tex}
\textbf{Ablation Study.} We ablate the \emph{Dynamic Semantic Trigger} and the \emph{Dual-Role Stream Orchestrator} to quantify their contributions. Table~\ref{tab:ablation_avg} summarizes the results averaged over all benchmarks, while Figure~\ref{fig:ablation_radar} visualizes the trade-offs across quality, responsiveness, reaction speed, computational efficiency, and interaction stability. Detailed per-benchmark results are reported in Appendix~\ref{app:ablation}.
Removing the trigger (\textbf{w/o Trigger}) reduces accuracy and substantially increases overhead, with NFE rising from 2.10 to 7.23 and the interruption rate from 8.20\% to 60.12\%. Removing the orchestrator (\textbf{w/o Orchestrator}) further degrades accuracy and increases the interruption rate, despite similar NFE, indicating that orchestration is important for maintaining stable streaming reasoning.
We further evaluate LTS-VoiceAgent with a larger Qwen3-32B backbone in Appendix~\ref{app:backbones}, and observe consistent gains in reasoning quality while preserving sub-second streaming responsiveness.

\section{Conclusion}
We presented LTS-VoiceAgent, a streaming voice-agent framework that separates \emph{when to think} from \emph{how to reason}. LTS-VoiceAgent uses a Dynamic Semantic Trigger and a dual-role orchestration design to support ``thinking while speaking'' under evolving ASR transcripts. Across VERA, Spoken-MQA, BigBenchAudio, and our Pause-and-Repair benchmark, LTS-VoiceAgent achieves a stronger quality--latency--efficiency trade-off than serial cascaded pipelines and existing streaming strategies.

\section*{Ethics Statement}

This work introduces a framework for low-latency, full-duplex voice interaction. While our primary goal is to enhance user experience in legitimate assistant scenarios, we acknowledge that the improved naturalness and responsiveness—particularly the ability to handle interruptions and generate near-instant replies—could potentially be misused to create more convincing automated systems for social engineering or unsolicited communication. Additionally, the "listening-while-thinking" mechanism implies continuous processing of audio streams; real-world deployments must strictly adhere to privacy protocols (e.g., local processing or explicit consent) to prevent the unintended analysis of background conversations. Finally, as our framework relies on general-purpose LLMs (e.g., Qwen) for reasoning, it inherits their potential limitations regarding hallucinations and biases, necessitating the use of safety guardrails in downstream applications.

\section*{Limitations}
Our evaluation covers a limited range of backbones, languages, and acoustic conditions. Although we adopt an audio-only protocol and introduce Pause-and-Repair to stress-test hesitations and self-corrections, the benchmarks still cannot capture the full diversity of real conversations, including multi-turn dialog dynamics, background noise, accents, domain-specific terminology, and user intent shifts. We also do not include human-subject studies on perceived responsiveness and usefulness, nor a comprehensive analysis of safety and privacy risks in deployment. Addressing these aspects will require broader data, stronger safeguards, and application-specific evaluation.

\bibliographystyle{acl_natbib}
\bibliography{custom}

\clearpage

\appendix

\section{Usage of LLM}
Large language models (ChatGPT) were used in two ways. First, they were used to improve the clarity and fluency of English writing. Second, they were used as a data-generation and annotation component in our synthetic data pipelines: (i) to insert special trigger tokens (e.g., “[T]”) for synthesizing training data for the Dynamic Semantic Trigger, and (ii) to rewrite/augment text benchmarks into scenario-based, disfluent spoken-style inputs (e.g., filler insertion and self-correction patterns) for constructing the Pause-and-Repair benchmark, following fixed prompts described in the paper/appendix. These models were not used to design the core method, choose evaluation metrics, or perform statistical analysis. All experimental results, comparisons, and interpretations were conducted by the authors, who take full responsibility for the content.

\section{Ablation Studies}
\label{app:ablation}

\input{table_ablation_acc_latency.tex}
\input{table_ablation_efficiency.tex}
\input{table_32b_acc_latency.tex}
\input{table_32b_efficiency.tex}

We conduct ablation studies to quantify the contribution of two core components in LTS-VoiceAgent: the Dynamic Semantic Trigger and the Dual-Role Stream Orchestrator. We evaluate two variants, one removing the trigger and falling back to mechanical triggering, and the other removing orchestration and running the pipeline without dual-role coordination. We report accuracy, latency, TTFS, and efficiency metrics on four benchmarks.

Removing the semantic trigger consistently degrades accuracy across benchmarks, with a pronounced drop on BigBenchAudio and VERA. Although mechanical triggering may yield slightly smaller latency or TTFS on some datasets, it substantially increases redundant computation and instability. Table~\ref{tab:efficiency_consolidated} shows that the trigger removal leads to much higher NFE and a markedly higher interruption rate, indicating frequent invalid reasoning attempts that are later interrupted by streaming transcript updates.

Disabling orchestration results in a larger loss of reasoning quality than removing the trigger on most benchmarks, while often reducing TTFS. This suggests that orchestration primarily contributes to preserving reasoning quality under evolving partial input rather than simply accelerating first responses. In addition, the interruption rate increases noticeably without orchestration, reflecting reduced interaction stability when background state maintenance and foreground response generation are not properly coordinated.

Overall, the full LTS-VoiceAgent achieves the strongest balance across accuracy, responsiveness, and efficiency. The semantic trigger is the main driver of computational efficiency and interruption robustness, while the orchestrator is critical for sustaining reasoning quality and stable streaming behavior.

\section{Robustness to Backbone Scaling}
\label{app:backbones}

To complement the main results with Qwen3-8B, we further evaluate LTS-VoiceAgent with a larger \textbf{Qwen3-32B} backbone. We reuse the same ASR/TTS stack and experimental protocol as in Section~\ref{sec:experiments}, and report accuracy, latency, TTFS, and efficiency metrics to assess whether LTS-VoiceAgent remains effective when scaling up the underlying model.

\paragraph{Qwen3-32B improves streaming reasoning quality.}
Table~\ref{tab:backbone_32b_main_results} shows that upgrading the backbone from Qwen3-8B to Qwen3-32B consistently increases accuracy across all benchmarks. The gains are particularly clear on Spoken-MQA and our Pause-and-Repair benchmark, indicating that a stronger backbone translates into better end-to-end reasoning even under streaming constraints.

\paragraph{LTS-VoiceAgent preserves responsiveness compared with serial cascades.}
Even with the larger 32B model, LTS-VoiceAgent maintains sub-second TTFS, while the serial cascade baseline with Qwen3-32B exhibits multi-second TTFS across benchmarks. This confirms that the proposed streaming design remains crucial for interactivity as model size increases.

\paragraph{Efficiency remains stable, with a modest increase in interruption rate.}
As shown in Table~\ref{tab:backbone_32b_efficiency}, LTS-VoiceAgent with Qwen3-32B keeps NFE close to two across benchmarks, suggesting that scaling does not inflate the number of reasoning invocations. At the same time, the interruption rate is slightly higher than with Qwen3-8B, consistent with the intuition that longer per-step inference can make intermediate thoughts more susceptible to being overtaken by subsequent transcript updates.

\section{Qualitative Case Analysis}
\label{sec:qualitative_analysis}

To evaluate the robustness of LTS-VoiceAgent against real-world streaming noise, we present a qualitative analysis of three representative scenarios: verbal self-correction, complex logical reasoning, and long-context transformation. As illustrated in Figure \ref{fig:case}, all inputs retain original ASR errors (e.g., transcription noise and segmentation faults) to stress-test the systems. We compare our method against baselines including Serial Processing, as well as existing streaming strategies such as VAD-based method and PredGen.

\subsection*{Frequent Verbal Self-Correction}
\begin{figure*}[p]
    \centering
    \includegraphics[width=1\linewidth, height=0.95\textheight, keepaspectratio]{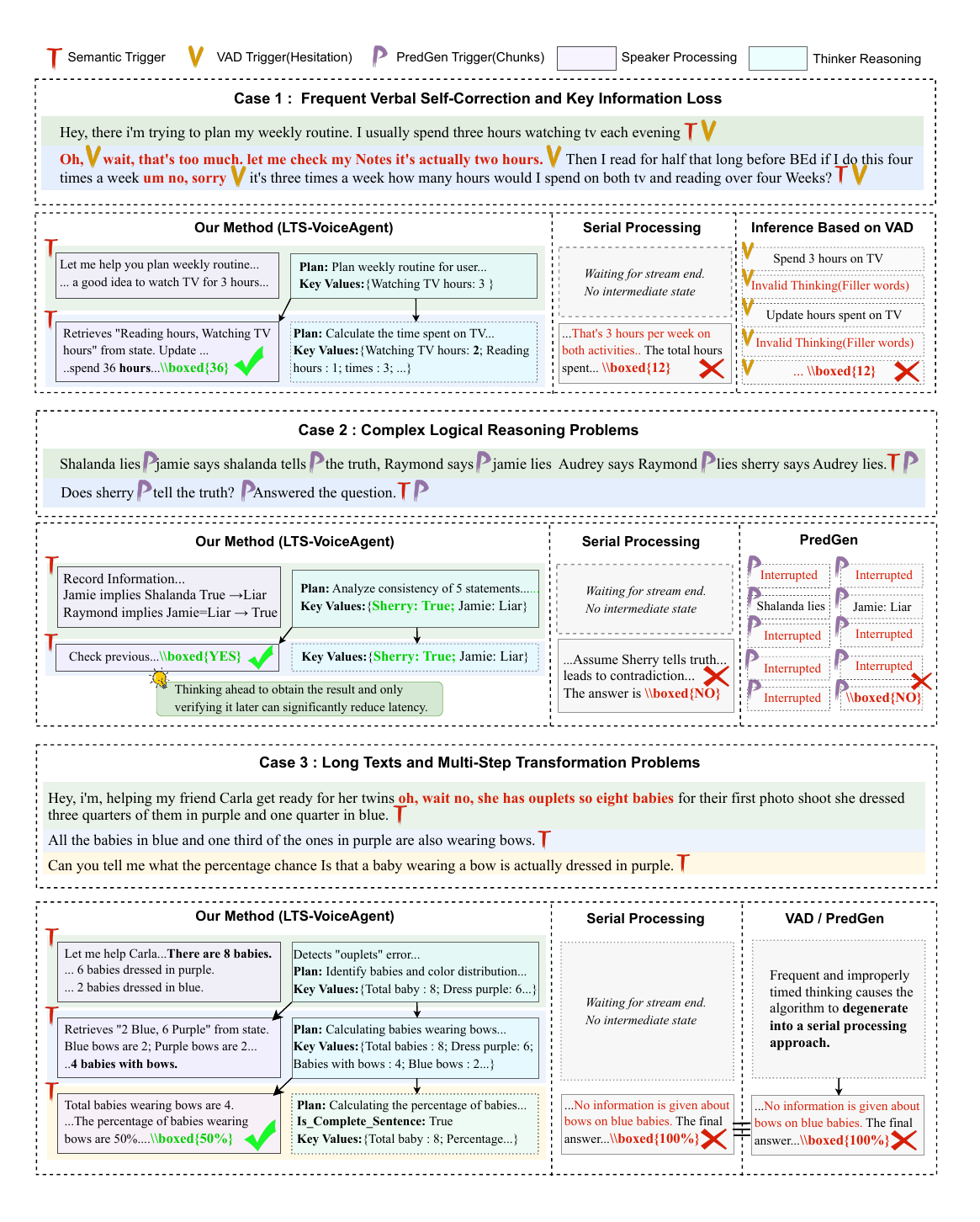}
    
    \caption{Qualitative comparison of LTS-VoiceAgent against baselines across three scenarios. \textbf{Top:} Handling verbal self-corrections via state overwriting. \textbf{Middle:} Solving logical paradoxes via look-ahead reasoning. \textbf{Bottom:} Managing long-context constraints via incremental updates. The inputs contain raw ASR errors to demonstrate robustness.}
    \label{fig:case}
\end{figure*}

As shown in the top panel of Figure \ref{fig:case}, the user verbally corrects the duration of a scheduled activity (modifying "3 hours" to "2 hours").

\textbf{LTS-VoiceAgent:} By maintaining a structured JSON state, the \textit{Thinker} detects the semantic correction and immediately overwrites the variable \texttt{Watching TV hours} from 3 to 2. This ensures the final calculation yields the correct result (36) based on the updated state.

\textbf{Baselines:} The Serial baseline, lacking intermediate state maintenance, fails to capture the overwrite semantics. It performs a linear pass after the stream ends, using the obsolete value to produce an incorrect result (12). Meanwhile, VAD-based method trigger premature reasoning on hesitation markers (e.g., "um no"), causing unnecessary intermediate reasoning and degrading consistency.

\input{table_dataset.tex}

\subsection*{Complex Logical Reasoning}
The middle panel illustrates a multi-hop logical paradox involving conflicting truthfulness statements (e.g., "Shalanda lies", "Jamie says Shalanda tells the truth").

\textbf{LTS-VoiceAgent:} The \textit{Thinker} constructs the logical chain in the background, anchoring key states (e.g., \texttt{Jamie: Liar}) while the user speaks. The system resolves the paradox incrementally and only requires a final verification step, avoiding dead loops and minimizing response latency.

\textbf{Baselines:} The Serial Processing relies on linear chain-of-thought processing. A single hallucination during the intermediate inference causes the entire logical verification to collapse, leading to an incorrect prediction. The PredGen strategy performs poorly here; frequent interruptions (250–350ms) fragment the reasoning process, preventing the formation of necessary deep logical dependencies.

\subsection*{Long-Context Multi-Step Transformation}
The bottom panel depicts a math word problem containing ASR noise (e.g., "octuplets" recognized as "ouplets") and multiple constraints.

\textbf{LTS-VoiceAgent:} Despite the noise, the \textit{Thinker} continuously updates state variables (e.g., \texttt{Blue bows: 2}) throughout the stream. This incremental maintenance ensures that early constraints (color distribution) are preserved, leading to the correct probability calculation (50\%).

\textbf{Baselines:} The Serial baseline suffers from forgetting; it fails to recall the specific constraints regarding blue babies mentioned early in the prompt, resulting in a logical gap and an incorrect answer (100\%). Similarly, VAD-based and PredGen methods degrade to serial-like performance as improper trigger timing disrupts semantic continuity.

\section{Dataset Construction Details}

\label{sec:appendix_data}
This appendix details the construction of our synthetic streaming speech dataset described in Section~\ref{sec:dataset}. We propose a pipeline to transform static text benchmarks into realistic spoken interactions, focusing on simulating complex disfluencies such as hesitation and self-correction.

\subsection{Data Processing Pipeline}
To bridge the gap between text-based benchmarks (GSM8K, MMLU-Pro) and real-world speech interaction, we implemented the multi-stage pipeline illustrated in Algorithm \ref{alg:data_pipeline}. The process begins with removing questions unsuitable for audio (e.g., extensive multiple-choice lists). The core transformation involves two phases:
\begin{enumerate}
    \item \textbf{Scenario-based Contextualization:} Abstract queries are rewritten into first-person, daily-life scenarios to ensure contextual relevance.
    \item \textbf{Hybrid Linguistic Perturbation:} We employ an LLM to simultaneously inject filler words and simulate cognitive backtracking (self-correction), creating a challenging "Hybrid" test set.
\end{enumerate}
Finally, the processed text is converted into audio using a TTS model.

\input{table_dataset_pipeline.tex}

\subsection{Prompt Engineering for Naturalness}
We utilized the \textbf{GPT-4o}  as the core augmentation engine. To ensure high-quality simulation of spontaneous speech, we designed specific system prompts that leverage the model's Chain-of-Thought (CoT) capabilities. 

\paragraph{Phase 1: Scenario-based Contextualization}
This phase transforms third-person exam questions into first-person intents. The prompt (see Figure~\ref{fig:system_prompts}, Top) enforces a "Role-Play" paradigm to ensure the query sounds like a natural request to a voice assistant.

\paragraph{Phase 2: Linguistic Perturbation}
To simulate spontaneous speech, we focus on the \textbf{Hybrid} setting, which combines hesitation markers with self-correction logic. This prompt (see Figure~\ref{fig:system_prompts}, Bottom) directs the LLM to identify "cognitive load points" for inserting fillers and "intent shifts" for corrections.

\subsection{Text-to-Speech Synthesis}
The linguistically perturbed text is synthesized using \textbf{CosyVoice 2}. We employ specific instructional prompts (e.g., \textit{"Pay attention to pauses, control speaking rate"}) to ensure the model renders the inserted text artifacts ("umm", "wait, no") as natural prosodic events rather than standard lexical tokens.

\subsection{Dataset Statistics}
Table \ref{tab:data_stats} presents the detailed statistics of our generated dataset. The dataset covers multiple domains from GSM8K (Mathematics) and MMLU-Pro (STEM, Humanities, etc.). We ensured a balanced distribution across the three augmentation types to robustly test the model's performance under different noise conditions.

\subsection{Licensing and Terms}
We release our Pause-and-Repair Benchmark under the MIT License, ensuring consistency with the source datasets. Regarding the existing artifacts utilized in this work, we strictly adhere to their original licenses: GSM8K and MMLU-Pro are used under the MIT License; BigBench Audio, Qwen3, and CosyVoice 2 are used under the Apache 2.0 License. Our use of these assets for academic research is consistent with their intended purposes and terms of use.

\begin{figure*}[p]
    \centering
    \input{prompt_scenario_context.tex}
    
    \vspace{0.4cm}  
    
    \input{prompt_hybrid.tex}
    
    \vspace{-0.2cm}
    \caption{Detailed system prompts for the two phases of data construction. Top: Phase 1 Scenario-based Contextualization; Bottom: Phase 2 Hybrid Perturbation.}
    \label{fig:system_prompts}
\end{figure*}

\end{document}

%% file: table_acc_latency.tex
\begin{table*}[t]
\centering
\small
\setlength{\tabcolsep}{3.5pt}
\renewcommand{\arraystretch}{1.15}
\resizebox{\textwidth}{!}{
\begin{tabular}{l ccc ccc ccc ccc}
\toprule
\multirow{2}{*}{\textbf{Method}} &
\multicolumn{3}{c}{\textbf{VERA}} &
\multicolumn{3}{c}{\textbf{Spoken-MQA}} &
\multicolumn{3}{c}{\textbf{BigBench}} &
\multicolumn{3}{c}{\textbf{Our Bench}} \\
\cmidrule(lr){2-4}\cmidrule(lr){5-7}\cmidrule(lr){8-10}\cmidrule(lr){11-13}
& \textbf{Acc}(\%) & \textbf{Lat} & \textbf{TTFS}
& \textbf{Acc}(\%) & \textbf{Lat} & \textbf{TTFS}
& \textbf{Acc}(\%) & \textbf{Lat} & \textbf{TTFS} 
& \textbf{Acc}(\%) & \textbf{Lat} & \textbf{TTFS} \\
\midrule
\textbf{Serial} (No-think)
  & 4.36 & 1205 & 4782   & 75.53 & 904 & 3557 & 54.95 & 990 & 3166 
  & 58.45 & 663 & 1764 \\
\textbf{Serial} (Think, upper bound)
  & 18.54 & 151k & 212k & 83.95 & 37k & 43k & 73.77 & 76k & 84k 
  & 76.34 & 51k & 61k \\
\addlinespace[3pt]
\textbf{PredGen} (No-think)
  & 4.34 & 247 & \textbf{298} & 67.87 & 230 & 281 & 54.5 & 213 & 264 
  & 55.84 & 251 & 303 \\
\textbf{PredGen} (Think)
  & 8.69 & 47k & 106k & 83.52 & 23k & 25k & 71.6 & 33k & 46k 
  & 70.51 & 43k & 48k \\
\addlinespace[3pt]
\textbf{LTS-VAD} (No-think)
  & 3.98 & 244 & 380 & 75.14 & 213 & \textbf{266} & 55.2 & 206 & \textbf{258} 
  & 56.41 & 244 & \textbf{294} \\
\textbf{LTS-VAD} (Think)
  & 8.69 & 47k & 103k & 83.52 & 24k & 26k & 71.5 & 33k & 45k 
  & 72.36 & 29k & 38k \\
\addlinespace[3pt]
\textbf{LTS-VoiceAgent (Ours)}
  & \textbf{6.88} & \textbf{230} & 415
  & \textbf{78.52} & 207 & 332
  & \textbf{71.9} & 209 & 338 
  & \textbf{62.57} & \textbf{236} & 336 \\
\addlinespace[3pt]
\textbf{Qwen2.5-Omni} (E2E)
  & 1.44 & 339.9 & 985.95 & 59.89 & \textbf{139.44} & 634.68 & 43.1 & \textbf{195.13} & 332.92
  & 50.25 & 245.17 & 449.9 \\
\bottomrule
\end{tabular}
}
\caption{Overall performance (Accuracy, Latency, and TTFS) across three public benchmarks and our proposed benchmark. Latency and TTFS are measured in milliseconds (ms). Serial (Think) provides an upper bound with full context; bold marks the best among real-time methods. Large values are rounded to 'k' for compactness.}
\label{tab:main_results_consolidated}
\end{table*}

%% file: table_efficiency.tex
\begin{table*}[t]
\centering
\small
\setlength{\tabcolsep}{3pt}
\renewcommand{\arraystretch}{1.15}
\resizebox{\textwidth}{!}{
\begin{tabular}{l ccc ccc ccc ccc}
\toprule
\multirow{2}{*}{\textbf{Method}} &
\multicolumn{3}{c}{\textbf{VERA}} &
\multicolumn{3}{c}{\textbf{Spoken-MQA}} &
\multicolumn{3}{c}{\textbf{BigBench}} &
\multicolumn{3}{c}{\textbf{Our Bench}} \\
\cmidrule(lr){2-4}\cmidrule(lr){5-7}\cmidrule(lr){8-10}\cmidrule(lr){11-13}
& \textbf{NFE}$\downarrow$ & \textbf{NIT}$\downarrow$ & \textbf{Rate}(\%)
& \textbf{NFE}$\downarrow$ & \textbf{NIT}$\downarrow$ & \textbf{Rate}(\%)
& \textbf{NFE}$\downarrow$ & \textbf{NIT}$\downarrow$ & \textbf{Rate}(\%)
& \textbf{NFE}$\downarrow$ & \textbf{NIT}$\downarrow$ & \textbf{Rate}(\%) \\
\midrule

\textbf{PredGen} (No-think)
  & 176.83 & 175.11 & 99.03
  & 59.39 & 58.14 & 97.90
  & 80.73 & 79.38 & 98.33
  & 172.46 & 156.06 & 90.05 \\
\textbf{PredGen} (Think)
  & 176.83 & 175.83 & 99.43
  & 59.39 & 58.39 & 98.31
  & 80.73 & 79.73 & 98.76
  & 172.46 & 164.34 & 95.29 \\

\addlinespace[3pt]
\textbf{LTS-VAD} (No-think)
  & 10.94 & 8.04 & 73.49
  & 3.42 & 1.82 & 53.21
  & 6.28 & 4.68 & 74.52
  & 8.23 & 3.42 & 41.53 \\
\textbf{LTS-VAD} (Think)
  & 10.94 & 9.80 & 89.58
  & 3.42 & 2.39 & 69.88
  & 6.29 & 5.26 & 83.62
  & 8.23 & 6.22 & 75.49 \\

\addlinespace[3pt]
\textbf{LTS-VoiceAgent (Ours)}
  & \textbf{2.37} & \textbf{0.13} & \textbf{5.46}
  & \textbf{1.96} & \textbf{0.16} & \textbf{8.16}
  & \textbf{2.05} & \textbf{0.20} & \textbf{9.76}
  & \textbf{2.02} & \textbf{0.19} & \textbf{9.4} \\

\bottomrule
\end{tabular}
}
\caption{Efficiency analysis comparing Number of Forward-pass Evaluations (NFE), Number of Interruption Times (NIT), and Interruption Rate across all datasets. Lower values ($\downarrow$) indicate better efficiency and fewer unnecessary interruptions. \textbf{Bold} indicates the best performance.}
\label{tab:efficiency_consolidated}
\end{table*}

%% file: table_ablation_avg.tex
\begin{table}[t]
\centering
\small
\setlength{\tabcolsep}{7pt}
\renewcommand{\arraystretch}{1.15}
\begin{tabular}{lccc}
\toprule
\textbf{Metric} & \textbf{Ours} & \textbf{w/o Trigger} & \textbf{w/o Orch.} \\
\midrule
Acc (\%)$\uparrow$        & 55.37 & 52.30 & 49.37 \\
Latency (ms)$\downarrow$  & 238.38 & 210.75 & 225.58 \\
TTFS (ms)$\downarrow$     & 339.59 & 332.22 & 282.40 \\
NFE$\downarrow$           & 2.10  & 7.23  & 2.10  \\
Int.Rate (\%)$\downarrow$ & 8.20  & 60.12 & 14.81 \\
\bottomrule
\end{tabular}
\caption{Ablation results averaged over all benchmarks. Int.Rate denotes the interruption rate (NIT/NFE). Detailed per-benchmark results are in Appendix~\ref{app:ablation}.}
\label{tab:ablation_avg}
\end{table}

%% file: table_ablation_acc_latency.tex
\begin{table*}[t]
\centering
\small
\setlength{\tabcolsep}{3.5pt}
\renewcommand{\arraystretch}{1.15}
\resizebox{\textwidth}{!}{
\begin{tabular}{l ccc ccc ccc ccc}
\toprule
\multirow{2}{*}{\textbf{Method}} &
\multicolumn{3}{c}{\textbf{VERA}} &
\multicolumn{3}{c}{\textbf{Spoken-MQA}} &
\multicolumn{3}{c}{\textbf{BigBench}} &
\multicolumn{3}{c}{\textbf{Our Bench}} \\
\cmidrule(lr){2-4}\cmidrule(lr){5-7}\cmidrule(lr){8-10}\cmidrule(lr){11-13}
& \textbf{Acc}(\%) & \textbf{Latency} & \textbf{TTFS}
& \textbf{Acc}(\%) & \textbf{Latency} & \textbf{TTFS}
& \textbf{Acc}(\%) & \textbf{Latency} & \textbf{TTFS} 
& \textbf{Acc}(\%) & \textbf{Latency} & \textbf{TTFS} \\
\midrule
\textbf{LTS-VoiceAgent w/o Trigger}
  & 5.79 & 224.36 & 349.62 & 76.91 & 183.22 & 298.71 & 66.1 & 195.51 & 309.37 
  & 60.41 & 239.92 & 371.16 \\
\textbf{LTS-VoiceAgent w/o orchestrator}
  & 3.26 & 236.99 & 288.69 & 67.96 & 219.32 & 272.64 & 64.5 & 211.38 & 263.65 
  & 61.77 & 234.64 & 304.60 \\
\textbf{LTS-VoiceAgent (Ours)}
  & 6.88 & 230.74 & 415.49
  & 78.52 & 207.46 & 332.48
  & 71.9 & 209.41 & 338.5
  & 62.57 & 235.97 & 335.95 \\

\bottomrule
\end{tabular}
}
\caption{Ablation results on accuracy, latency, and TTFS across three public benchmarks and our Pause-and-Repair benchmark. Latency and TTFS are measured in milliseconds (ms).}
\label{tab:table_ablation_acc_latency}
\end{table*}

%% file: table_ablation_efficiency.tex
\begin{table*}[t]
\centering
\small
\setlength{\tabcolsep}{3pt}
\renewcommand{\arraystretch}{1.15}
\resizebox{\textwidth}{!}{
\begin{tabular}{l ccc ccc ccc ccc}
\toprule
\multirow{2}{*}{\textbf{Method}} &
\multicolumn{3}{c}{\textbf{VERA}} &
\multicolumn{3}{c}{\textbf{Spoken-MQA}} &
\multicolumn{3}{c}{\textbf{BigBench}} &
\multicolumn{3}{c}{\textbf{Our Bench}} \\
\cmidrule(lr){2-4}\cmidrule(lr){5-7}\cmidrule(lr){8-10}\cmidrule(lr){11-13}
& \textbf{NFE}$\downarrow$ & \textbf{NIT}$\downarrow$ & \textbf{Rate}(\%)
& \textbf{NFE}$\downarrow$ & \textbf{NIT}$\downarrow$ & \textbf{Rate}(\%)
& \textbf{NFE}$\downarrow$ & \textbf{NIT}$\downarrow$ & \textbf{Rate}(\%)
& \textbf{NFE}$\downarrow$ & \textbf{NIT}$\downarrow$ & \textbf{Rate}(\%) \\
\midrule

\textbf{LTS-VoiceAgent w/o Trigger}
  & 10.94 & 7.64 & 69.64
  & 3.42 & 1.94 & 56.73
  & 6.28 & 3.54 & 56.37
  & 8.23 & 4.76 & 57.75 \\
\textbf{LTS-VoiceAgent w/o orchestrator}
  & 2.37 & 0.34 & 14.00
  & 1.96 & 0.12 & 6.12
  & 2.05 & 0.35 & 17.07
  & 2.02 & 0.45 & 22.05 \\
\textbf{LTS-VoiceAgent (Ours)}
  & 2.37 & 0.13 & 5.46
  & 1.96 & 0.16 & 8.16
  & 2.05 & 0.20 & 9.76
  & 2.02 & 0.19 & 9.4 \\

\bottomrule
\end{tabular}
}
\caption{Efficiency analysis for ablations, reporting NFE, NIT, and interruption rate (NIT/NFE). Lower values indicate fewer redundant evaluations and more stable streaming interaction.}
\label{tab:ablation_efficiency}
\end{table*}

%% file: table_32b_acc_latency.tex
\begin{table*}[t!]
\centering
\small
\setlength{\tabcolsep}{3.5pt}
\renewcommand{\arraystretch}{1.15}
\resizebox{\textwidth}{!}{
\begin{tabular}{l ccc ccc ccc ccc}
\toprule
\multirow{2}{*}{\textbf{Method}} &
\multicolumn{3}{c}{\textbf{VERA}} &
\multicolumn{3}{c}{\textbf{Spoken-MQA}} &
\multicolumn{3}{c}{\textbf{BigBench}} &
\multicolumn{3}{c}{\textbf{Our Bench}} \\
\cmidrule(lr){2-4}\cmidrule(lr){5-7}\cmidrule(lr){8-10}\cmidrule(lr){11-13}
& \textbf{Acc}(\%) & \textbf{Latency} & \textbf{TTFS}
& \textbf{Acc}(\%) & \textbf{Latency} & \textbf{TTFS}
& \textbf{Acc}(\%) & \textbf{Latency} & \textbf{TTFS} 
& \textbf{Acc}(\%) & \textbf{Latency} & \textbf{TTFS} \\
\midrule
\textbf{Qwen3-32B} (Serial)
  & 5.45 & 726.35 & 3929.33 & 79.89 & 708.4 & 2588.86 & 63.26 & 711.01 & 2367.32
  & 66.84 & 723.66 & 2262.58 \\
\textbf{Qwen3-32B} (LTS-VoiceAgent)
  & 8.15 & 377.47 & 402.78
  & 85.73 & 309.79 & 592.22
  & 73.4 & 301.88 & 588.92
  & 73.89 & 357.44 & 829.98 \\

\bottomrule
\end{tabular}
}
\caption{Performance of Qwen3-32B under a serial cascade and with LTS-VoiceAgent across all benchmarks. Latency and TTFS are measured in milliseconds (ms).}
\label{tab:backbone_32b_main_results}
\end{table*}

%% file: table_32b_efficiency.tex
\begin{table*}[t!]
\centering
\small
\setlength{\tabcolsep}{3pt}
\renewcommand{\arraystretch}{1.15}
\resizebox{\textwidth}{!}{
\begin{tabular}{l ccc ccc ccc ccc}
\toprule
\multirow{2}{*}{\textbf{Method}} &
\multicolumn{3}{c}{\textbf{VERA}} &
\multicolumn{3}{c}{\textbf{Spoken-MQA}} &
\multicolumn{3}{c}{\textbf{BigBench}} &
\multicolumn{3}{c}{\textbf{Our Bench}} \\
\cmidrule(lr){2-4}\cmidrule(lr){5-7}\cmidrule(lr){8-10}\cmidrule(lr){11-13}
& \textbf{NFE}$\downarrow$ & \textbf{NIT}$\downarrow$ & \textbf{Rate}(\%)
& \textbf{NFE}$\downarrow$ & \textbf{NIT}$\downarrow$ & \textbf{Rate}(\%)
& \textbf{NFE}$\downarrow$ & \textbf{NIT}$\downarrow$ & \textbf{Rate}(\%)
& \textbf{NFE}$\downarrow$ & \textbf{NIT}$\downarrow$ & \textbf{Rate}(\%) \\
\midrule

\textbf{Qwen3-32B} (LTS-VoiceAgent)
  & 2.37 & 0.33 & 13.9
  & 1.96 & 0.28 & 14.28
  & 2.05 & 0.23 & 11.2
  & 2.02 & 0.20 & 10.05 \\

\bottomrule
\end{tabular}%
}
\caption{Efficiency of LTS-VoiceAgent with Qwen3-32B, reporting NFE, NIT, and interruption rate (NIT/NFE) across all benchmarks. Lower values indicate better efficiency and fewer interruptions.}
\label{tab:backbone_32b_efficiency}
\end{table*}

%% file: table_dataset.tex
\begin{table*}[t!]
\centering
\small
\resizebox{\textwidth}{!}{
\begin{tabular}{@{}llcccc@{}}
\toprule
\textbf{Source} & \textbf{Category/Domain} & \textbf{Hybrid Samples} & \textbf{Avg. Tokens} & \textbf{Avg. Audio (s)} & \textbf{Vocab Diversity} \\ \midrule

\textbf{GSM8K} & Mathematics  & \textbf{1319} & \textbf{67.6} & \textbf{25.2} & Med \\ \midrule

\multirow{7}{*}{\textbf{MMLU-Pro}} 
& Math  & 111 & 104.7 & 50.1 & Hard \\
& Physics  & 104 & 121.4 & 57.2 & Hard \\
& Chemistry  & 92 & 114.1 & 60.5 & Hard \\
& Law  & 89 & 179.9 & 71.7 & Hard \\
& Engineering  & 80 & 142.5 & 77.5 & Med \\
& Others (Bio, Chem, etc.)  & 507 & 113.3 & 52.2 & Med \\
& \textit{MMLU Subtotal} & \textbf{983} & \textbf{121.7} & \textbf{57.1} & \textbf{Hard} \\ 

\midrule 

\textbf{Total} & \textbf{All Domains} & \textbf{2302} & \textbf{90.7} & \textbf{38.9} & \textbf{-} \\ \bottomrule
\end{tabular}
}
\caption{Detailed statistics of the synthetic dataset. "Avg. Tokens" refers to the text length after augmentation, and "Avg. Audio" estimates the duration of the synthesized speech. The sample counts for MMLU-Pro subsets are adjusted to ensure a balanced comparison.}
\label{tab:data_stats}
\end{table*}

%% file: table_dataset_pipeline.tex
\begin{algorithm}[t]
\caption{Synthetic Spoken Dataset Construction Pipeline}
\label{alg:data_pipeline}
\small
\begin{algorithmic}[1]
\Require Datasets $\mathcal{D}_{src} = \{GSM8K, MMLU\text{-}Pro\}$
\Require TTS Model $\mathcal{M}_{tts}$ (CosyVoice 2)
\Ensure Augmented Audio Dataset $\mathcal{A}_{hybrid}$

\State $\mathcal{D}_{clean} \gets \emptyset$
\Statex \textcolor{blue}{\it \# Phase 1: Scenario-based Contextualization}
\For{each sample $s \in \mathcal{D}_{src}$}
    \If{$s$ is suitable} 
        \State $s' \gets \Call{ScenarioWrapper}{s}$ 
        \State $\mathcal{D}_{clean}.\Call{append}{s'}$
    \EndIf
\EndFor

\Statex \textcolor{blue}{\it \# Phase 2: Hybrid Linguistic Perturbation}
\State $\mathcal{T}_{hybrid} \gets \Call{LLM\_Augment}{\mathcal{D}_{clean}, \text{Prompt}_{hybrid}}$

\Statex \textcolor{blue}{\it \# Phase 3: TTS Synthesis}
\State $\mathcal{P}_{audio} \gets \text{"Note pauses in tone, control speed..."}$
\For{$text \in \mathcal{T}_{hybrid}$}
    \State $audio \gets \mathcal{M}_{tts}(text, \text{prompt}=\mathcal{P}_{audio})$
    \State $\mathcal{A}_{hybrid}.\Call{append}{audio}$
\EndFor

\State \Return $\mathcal{A}_{hybrid}$
\end{algorithmic}
\end{algorithm}

%% file: prompt_scenario_context.tex
\begin{promptbox}[System Prompt: Scenario-based Contextualization]
    \textbf{Role:} You are an expert conversational designer and creative writer for voice AI assistants. Your task is to rewrite standard math word problems into natural, first-person spoken queries addressed to a virtual assistant.

    \vspace{0.2cm}

    \textbf{Task:} Rewrite math word problems into natural, first-person spoken queries.

    \vspace{0.2cm}

    \textbf{Core Guidelines:}
    
    1. First-Person Perspective: Change all third-person narratives to first-person ("I", "me", "my"). If the original problem has no people, inject yourself as the protagonist.

    \vspace{0.1cm}

    2. Real-Life Context: Construct a realistic daily life scenario (e.g., budgeting, cooking, helping a child with homework).

    \vspace{0.1cm}

    3. Conversational Openers: Start every rewritten question with a natural spoken greeting (e.g., "Hey," "Hi there," "Ok computer") and a brief intro phrase.

    \vspace{0.1cm}

    4. Preserve Logic: The mathematical values and logic must remain exactly the same.

    \vspace{0.1cm}

    5. Natural Flow: The output should sound like something a human would actually say aloud.

    \vspace{0.2cm}
    \hrule
    \vspace{0.2cm}
    
    \textbf{Example (Condensed):}
    
    \textit{Input:} "A bag has 232 candies. 54 are red. How many are pink?"
    
    \textit{Output:} "Hi there. I bought a huge bag of Starbursts with 232 pieces for the party... I only know there are 54 red ones... Could you tell me how many pink candies are left?"
\end{promptbox}

%% file: prompt_hybrid.tex
\begin{promptbox}[System Prompt: Hybrid Perturbation]
    \textbf{Role:}You are a data augmentation expert simulating natural, spontaneous human speech. Rewrite the input text to make it sound like a real person talking by inserting \textbf{Filler Words} and simulating \textbf{Self-Corrections} (changing mind/fixing errors).

    \vspace{0.2cm}

    \textbf{ Critical Rules:} 
    
    1. No Solving: Do NOT answer questions. Only rewrite the text provided.

    \vspace{0.1cm}

    2. Structure Preservation:Keep labels like "Person A:", "Q:", "Answer:" exactly as they are.
If the input contains a list of options (e.g., A, B, C...), DO NOT modify the options. Only insert fillers/corrections into the main question.

    \vspace{0.1cm}

    3. Density Limits:Mix fillers and corrections naturally. Do not overdo it.
Placement: Place fillers at logical pauses; place corrections where a logical thought change could occur.
    
    \vspace{0.2cm}
        
    \textbf{Filler Words Vocabulary:}
    \begin{itemize}[nosep]
        \item \textit{Hesitation:} "umm", "ah", "let me see", "basically"
        \item \textit{Correction:} "wait, no", "actually, scrap that", "I mean"
        \item \textit{Delaying:} "uh", "hmm", "so", "anyway"
        \item \textit{Transition:} "well", "like", "you know", "actually", "basically"
    \end{itemize}

    \vspace{0.2cm}

    \textbf{Self-Correction Patterns:}
    \begin{itemize}[nosep]
        \item \textit{Entity Swap:} "In 2015... wait, 2017."
        \item \textit{Misreading:} "The man walked... oh, the woman walked."
        \item \textit{Logic Check:} "So I should agree... actually, thinking about the consequences, I disagree."
        \item \textit{Distraction:} "I need to buy apples... thinking of pie... anyway, I need apples."
    \end{itemize}

    \vspace{0.2cm}
    \hrule
    \vspace{0.2cm}
    
    \textbf{Example (Condensed):}
    
    \textit{Input:} "I want to set the price low to gain market share. What is this strategy called?"
    
    \textit{Output:} "I want to set the price high... \textbf{wait, no}, I want to set the price low to \textbf{umm}, let me see, gain market share. What is this strategy called?"
\end{promptbox}

%% file: custom.bib
@article{hurst2024gpt,
  title   = {Gpt-4o system card},
  author  = {Hurst, Aaron and Lerer, Adam and Goucher, Adam P and Perelman, Adam and Ramesh, Aditya and Clark, Aidan and Ostrow, AJ and Welihinda, Akila and Hayes, Alan and Radford, Alec and others},
  journal = {arXiv preprint arXiv:2410.21276},
  year    = {2024}
}

@article{fang2024llama,
  title   = {Llama-omni: Seamless speech interaction with large language models},
  author  = {Fang, Qingkai and Guo, Shoutao and Zhou, Yan and Ma, Zhengrui and Zhang, Shaolei and Feng, Yang},
  journal = {arXiv preprint arXiv:2409.06666},
  year    = {2024}
}

@article{defossez2024moshi,
  title   = {Moshi: a speech-text foundation model for real-time dialogue},
  author  = {D{\'e}fossez, Alexandre and Mazar{\'e}, Laurent and Orsini, Manu and Royer, Am{\'e}lie and P{\'e}rez, Patrick and J{\'e}gou, Herv{\'e} and Grave, Edouard and Zeghidour, Neil},
  journal = {arXiv preprint arXiv:2410.00037},
  year    = {2024}
}

@article{li2025predgen,
  title   = {PredGen: Accelerated Inference of Large Language Models through Input-Time Speculation for Real-Time Speech Interaction},
  author  = {Li, Shufan and Grover, Aditya},
  journal = {arXiv preprint arXiv:2506.15556},
  year    = {2025}
}

@article{chen2024livemind,
  title   = {Livemind: Low-latency large language models with simultaneous inference},
  author  = {Chen, Chuangtao and Zhang, Grace Li and Yin, Xunzhao and Zhuo, Cheng and Schlichtmann, Ulf and Li, Bing},
  journal = {arXiv preprint arXiv:2406.14319},
  year    = {2024}
}

@inproceedings{ren2020simulspeech,
  title     = {SimulSpeech: End-to-end simultaneous speech to text translation},
  author    = {Ren, Yi and Liu, Jinglin and Tan, Xu and Zhang, Chen and Qin, Tao and Zhao, Zhou and Liu, Tie-Yan},
  booktitle = {Proceedings of the 58th Annual Meeting of the Association for Computational Linguistics},
  pages     = {3787--3796},
  year      = {2020}
}

@article{fang2025llama,
  title   = {Llama-omni2: Llm-based real-time spoken chatbot with autoregressive streaming speech synthesis},
  author  = {Fang, Qingkai and Zhou, Yan and Guo, Shoutao and Zhang, Shaolei and Feng, Yang},
  journal = {arXiv preprint arXiv:2505.02625},
  year    = {2025}
}

@article{wang2024freeze,
  title   = {Freeze-omni: A smart and low latency speech-to-speech dialogue model with frozen llm},
  author  = {Wang, Xiong and Li, Yangze and Fu, Chaoyou and Shen, Yunhang and Xie, Lei and Li, Ke and Sun, Xing and Ma, Long},
  journal = {arXiv preprint arXiv:2411.00774},
  year    = {2024}
}

@article{zhang2024intrinsicvoice,
  title   = {Intrinsicvoice: Empowering llms with intrinsic real-time voice interaction abilities},
  author  = {Zhang, Xin and Lyu, Xiang and Du, Zhihao and Chen, Qian and Zhang, Dong and Hu, Hangrui and Tan, Chaohong and Zhao, Tianyu and Wang, Yuxuan and Zhang, Bin and others},
  journal = {arXiv preprint arXiv:2410.08035},
  year    = {2024}
}

@article{long2025vita,
  title   = {VITA-Audio: Fast Interleaved Cross-Modal Token Generation for Efficient Large Speech-Language Model},
  author  = {Long, Zuwei and Shen, Yunhang and Fu, Chaoyou and Gao, Heting and Li, Lijiang and Chen, Peixian and Zhang, Mengdan and Shao, Hang and Li, Jian and Peng, Jinlong and others},
  journal = {arXiv preprint arXiv:2505.03739},
  year    = {2025}
}

@article{shi2025voila,
  title   = {Voila: Voice-Language Foundation Models for Real-Time Autonomous Interaction and Voice Role-Play},
  author  = {Shi, Yemin and Shu, Yu and Dong, Siwei and Liu, Guangyi and Sesay, Jaward and Li, Jingwen and Hu, Zhiting},
  journal = {arXiv preprint arXiv:2505.02707},
  year    = {2025}
}

@inproceedings{wei2025specasr,
  title        = {SpecASR: Accelerating LLM-based Automatic Speech Recognition via Speculative Decoding},
  author       = {Wei, Linye and Zhong, Shuzhang and Xu, Songqiang and Wang, Runsheng and Huang, Ru and Li, Meng},
  booktitle    = {2025 62nd ACM/IEEE Design Automation Conference (DAC)},
  pages        = {1--7},
  year         = {2025},
  organization = {IEEE}
}

@article{du2024cosyvoice,
  title   = {Cosyvoice 2: Scalable streaming speech synthesis with large language models},
  author  = {Du, Zhihao and Wang, Yuxuan and Chen, Qian and Shi, Xian and Lv, Xiang and Zhao, Tianyu and Gao, Zhifu and Yang, Yexin and Gao, Changfeng and Wang, Hui and others},
  journal = {arXiv preprint arXiv:2412.10117},
  year    = {2024}
}

@inproceedings{shikhar2025llmvox,
  title     = {LLMVoX: Autoregressive streaming text-to-speech model for any LLM},
  author    = {Shikhar, Sambal and Kurpath, Mohammed Irfan and Mullappilly, Sahal Shaji and Lahoud, Jean and Khan, Fahad Shahbaz and Anwer, Rao Muhammad and Khan, Salman and Cholakkal, Hisham},
  booktitle = {Findings of the Association for Computational Linguistics: ACL 2025},
  pages     = {20481--20493},
  year      = {2025}
}

@article{li2024style,
  title   = {Style-talker: Finetuning audio language model and style-based text-to-speech model for fast spoken dialogue generation},
  author  = {Li, Yinghao Aaron and Jiang, Xilin and Darefsky, Jordan and Zhu, Ge and Mesgarani, Nima},
  journal = {arXiv preprint arXiv:2408.11849},
  year    = {2024}
}

@article{likhomanenko2025chipchat,
  title   = {Chipchat: Low-latency cascaded conversational agent in mlx},
  author  = {Likhomanenko, Tatiana and Carlson, Luke and Bai, Richard He and Gu, Zijin and Tran, Han and Aldeneh, Zakaria and Zhang, Yizhe and Zhang, Ruixiang and Zheng, Huangjie and Jaitly, Navdeep},
  journal = {arXiv preprint arXiv:2509.00078},
  year    = {2025}
}

@article{ethiraj2025toward,
  title   = {Toward Low-Latency End-to-End Voice Agents for Telecommunications Using Streaming ASR, Quantized LLMs, and Real-Time TTS},
  author  = {Ethiraj, Vignesh and David, Ashwath and Menon, Sidhanth and Vijay, Divya},
  journal = {arXiv preprint arXiv:2508.04721},
  year    = {2025}
}

@article{goel2025audio,
  title   = {Audio flamingo 3: Advancing audio intelligence with fully open large audio language models},
  author  = {Goel, Arushi and Ghosh, Sreyan and Kim, Jaehyeon and Kumar, Sonal and Kong, Zhifeng and Lee, Sang-gil and Yang, Chao-Han Huck and Duraiswami, Ramani and Manocha, Dinesh and Valle, Rafael and others},
  journal = {arXiv preprint arXiv:2507.08128},
  year    = {2025}
}

@inproceedings{woo2025think,
  title     = {Think, Verbalize, then Speak: Bridging Complex Thoughts and Comprehensible Speech},
  author    = {Woo, Tony and Lee, Sehun and Kim, Kang-wook and Kim, Gunhee},
  booktitle = {Proceedings of the 2025 Conference on Empirical Methods in Natural Language Processing},
  pages     = {14373--14390},
  year      = {2025}
}

@article{cheng2024compressed,
  title   = {Compressed chain of thought: Efficient reasoning through dense representations},
  author  = {Cheng, Jeffrey and Van Durme, Benjamin},
  journal = {arXiv preprint arXiv:2412.13171},
  year    = {2024}
}

@article{hou2025thinkprune,
  title   = {Thinkprune: Pruning long chain-of-thought of llms via reinforcement learning},
  author  = {Hou, Bairu and Zhang, Yang and Ji, Jiabao and Liu, Yujian and Qian, Kaizhi and Andreas, Jacob and Chang, Shiyu},
  journal = {arXiv preprint arXiv:2504.01296},
  year    = {2025}
}

@article{xu2025chain,
  title   = {Chain of draft: Thinking faster by writing less},
  author  = {Xu, Silei and Xie, Wenhao and Zhao, Lingxiao and He, Pengcheng},
  journal = {arXiv preprint arXiv:2502.18600},
  year    = {2025}
}

@inproceedings{huang2024audiogpt,
  title     = {Audiogpt: Understanding and generating speech, music, sound, and talking head},
  author    = {Huang, Rongjie and Li, Mingze and Yang, Dongchao and Shi, Jiatong and Chang, Xuankai and Ye, Zhenhui and Wu, Yuning and Hong, Zhiqing and Huang, Jiawei and Liu, Jinglin and others},
  booktitle = {Proceedings of the AAAI Conference on Artificial Intelligence},
  volume    = {38},
  pages     = {23802--23804},
  year      = {2024}
}

@article{skantze2021turn,
  title     = {Turn-taking in conversational systems and human-robot interaction: a review},
  author    = {Skantze, Gabriel},
  journal   = {Computer Speech \& Language},
  volume    = {67},
  pages     = {101178},
  year      = {2021},
  publisher = {Elsevier}
}

@article{zhang2023speechgpt,
  title   = {Speechgpt: Empowering large language models with intrinsic cross-modal conversational abilities},
  author  = {Zhang, Dong and Li, Shimin and Zhang, Xin and Zhan, Jun and Wang, Pengyu and Zhou, Yaqian and Qiu, Xipeng},
  journal = {arXiv preprint arXiv:2305.11000},
  year    = {2023}
}

@article{chen2024voicebench,
  title   = {Voicebench: Benchmarking llm-based voice assistants},
  author  = {Chen, Yiming and Yue, Xianghu and Zhang, Chen and Gao, Xiaoxue and Tan, Robby T and Li, Haizhou},
  journal = {arXiv preprint arXiv:2410.17196},
  year    = {2024}
}

@article{jain2025voiceagentbench,
  title   = {VoiceAgentBench: Are Voice Assistants ready for agentic tasks?},
  author  = {Jain, Dhruv and Shukla, Harshit and Rajeev, Gautam and Kulkarni, Ashish and Khatri, Chandra and Agarwal, Shubham},
  journal = {arXiv preprint arXiv:2510.07978},
  year    = {2025}
}

@article{levinson2016turn,
  title     = {Turn-taking in human communication--origins and implications for language processing},
  author    = {Levinson, Stephen C},
  journal   = {Trends in cognitive sciences},
  volume    = {20},
  number    = {1},
  pages     = {6--14},
  year      = {2016},
  publisher = {Elsevier}
}

@article{lin2025voice,
  title   = {Voice evaluation of reasoning ability: Diagnosing the modality-induced performance gap},
  author  = {Lin, Yueqian and Hu, Zhengmian and Wang, Qinsi and Liu, Yudong and Zhang, Hengfan and Subramanian, Jayakumar and Vlassis, Nikos and Li, Hai Helen and Chen, Yiran},
  journal = {arXiv preprint arXiv:2509.26542},
  year    = {2025}
}

@article{wei2025towards,
  title   = {Towards Spoken Mathematical Reasoning: Benchmarking Speech-based Models over Multi-faceted Math Problems},
  author  = {Wei, Chengwei and Wang, Bin and Kim, Jung-jae and Chen, Nancy F},
  journal = {arXiv preprint arXiv:2505.15000},
  year    = {2025}
}

@article{suzgun2022challenging,
  title   = {Challenging BIG-Bench Tasks and Whether Chain-of-Thought Can Solve Them},
  author  = {Suzgun, Mirac and Scales, Nathan and Sch{\"a}rli, Nathanael and Gehrmann, Sebastian and Tay, Yi and Chung, Hyung Won and Chowdhery, Aakanksha and Le, Quoc V and Chi, Ed H and Zhou, Denny and Wei, Jason},
  journal = {arXiv preprint arXiv:2210.09261},
  year    = {2022}
}

@article{cobbe2021training,
  title   = {Training Verifiers to Solve Math Word Problems},
  author  = {Cobbe, Karl and Kosaraju, Vineet and Bavarian, Mohammad and Chen, Mark and Jun, Heewoo and Kaiser, Lukasz and Plappert, Matthias and Tworek, Jerry and Hilton, Jacob and Nakano, Reiichiro and others},
  journal = {arXiv preprint arXiv:2110.14168},
  year    = {2021}
}

@inproceedings{hendrycks2020measuring,
  title     = {Measuring Massive Multitask Language Understanding},
  author    = {Hendrycks, Dan and Burns, Collin and Basart, Steven and Zou, Andy and Mazeika, Mantas and Song, Dawn and Steinhardt, Jacob},
  booktitle = {Proceedings of the International Conference on Learning Representations (ICLR)},
  year      = {2021}
}

@article{wang2024mmlu,
  title   = {MMLU-Pro: A More Robust and Challenging Multi-Task Language Understanding Benchmark},
  author  = {Wang, Yubo and Ma, Xueguang and Zhang, Ge and Ni, Yuansheng and Chandra, Abhranil and Guo, Shiguang and Ren, Weiming and Arulraj, Aaran and He, Xuan and others},
  journal = {arXiv preprint arXiv:2406.01574},
  year    = {2024}
}

@article{brown2020language,
  title   = {Language models are few-shot learners},
  author  = {Brown, Tom and Mann, Benjamin and Ryder, Nick and Subbiah, Melanie and Kaplan, Jared D and Dhariwal, Prafulla and Neelakantan, Arvind and Shyam, Pranav and Sastry, Girish and Askell, Amanda and others},
  journal = {Advances in neural information processing systems},
  volume  = {33},
  pages   = {1877--1901},
  year    = {2020}
}

@article{achiam2023gpt,
  title   = {Gpt-4 technical report},
  author  = {Achiam, Josh and Adler, Steven and Agarwal, Sandhini and Ahmad, Lama and Akkaya, Ilge and Aleman, Florencia Leoni and Almeida, Diogo and Altenschmidt, Janko and Altman, Sam and Anadkat, Shyamal and others},
  journal = {arXiv preprint arXiv:2303.08774},
  year    = {2023}
}

@article{wei2022chain,
  title   = {Chain-of-thought prompting elicits reasoning in large language models},
  author  = {Wei, Jason and Wang, Xuezhi and Schuurmans, Dale and Bosma, Maarten and Xia, Fei and Chi, Ed and Le, Quoc V and Zhou, Denny and others},
  journal = {Advances in neural information processing systems},
  volume  = {35},
  pages   = {24824--24837},
  year    = {2022}
}

@inproceedings{radford2023robust,
  title        = {Robust speech recognition via large-scale weak supervision},
  author       = {Radford, Alec and Kim, Jong Wook and Xu, Tao and Brockman, Greg and McLeavey, Christine and Sutskever, Ilya},
  booktitle    = {International conference on machine learning},
  pages        = {28492--28518},
  year         = {2023},
  organization = {PMLR}
}

@article{wang2023neural,
  title   = {Neural codec language models are zero-shot text to speech synthesizers},
  author  = {Wang, Chengyi and Chen, Sanyuan and Wu, Yu and Zhang, Ziqiang and Zhou, Long and Liu, Shujie and Chen, Zhuo and Liu, Yanqing and Wang, Huaming and Li, Jinyu and others},
  journal = {arXiv preprint arXiv:2301.02111},
  year    = {2023}
}

@article{xu2025qwen2,
  title   = {Qwen2. 5-omni technical report},
  author  = {Xu, Jin and Guo, Zhifang and He, Jinzheng and Hu, Hangrui and He, Ting and Bai, Shuai and Chen, Keqin and Wang, Jialin and Fan, Yang and Dang, Kai and others},
  journal = {arXiv preprint arXiv:2503.20215},
  year    = {2025}
}
